# Measuring Race in US Economic Statistics: What Do We Know?[*]

Sonya Ravindranath Waddell, John M. Abowd, Camille Busette, and Mark Hugo Lopez

SONYA RAVINDRANATH WADDELL:

My name is Sonya Ravindranath Waddell. I'm here from the Federal Reserve Bank of Richmond, just somewhere between 90 minutes and five hours down 95. I've spent the better part of 20 years trying to understand our region and our world through data. This panel is going to help us understand how we measure race and ethnicity in the U.S. and what that means for how we understand economic outcomes for minority and immigrant populations in the country.

As we all know, and as we talked about this morning, fostering economic growth requires enabling all people, all Americans, to contribute fully to economic activity. When we don't enable our individuals, our households, and our communities to fully engage in the economy, we leave gains on the table.

But in order to address something with policy, we have to be able to understand it. In my world, and I think in the world of a lot of people in this room, in order to understand something, you have to be able to measure it. I'm hoping that we can all come out of this session with a sense of how we measure race, ethnicity, country of birth, *et cetera*; what, as researchers, we need to either understand or control for, at the very least in our minds, if not also in our models. And then what we, as data users, might be able to do or advocate for if we're seeking the best economic data that we can find.

We are now going to open up with questions for the panel. My first question is for John Abowd. John, what are the rules that govern how the Census and other agencies measure race and ethnicity?

JOHN ABOWD:

Thanks Sonya. Thank you to all of you for attending this session. Thank you, NABE, for organizing this conference and inviting me. I knew the first question was coming, and it deserves a thorough answer.

I asked a group of demographers at the Census Bureau to review this answer, and I want to acknowledge them here: Karen Battle, Nicholas Jones, and Rachel Marks. They are the true experts, and my role is to put their expertise in the context of economic measurement. Census Bureau demographers have prepared a very careful analysis comparing the 2010 and 2020 census redistricting data on race and ethnicity.

This research has shown that the best way to measure race and ethnicity is to use a single question that does not force the respondent to self-identify into categories "ethnicity" and "race." I'll have more to say about that later.

First, I need to explain the way race and ethnicity data are collected and disseminated, especially by statistical agencies, but in general in publications of the federal government. This process is not governed by the Census Bureau. It is the domain of the Office of Management and Budget's Statistical and Science Policy Branch in the Office of Information and Regulatory Affairs, led by the Chief Statistician of the United States. Most of these policies are issued as Statistical Policy Directives, and, in particular, Statistical Policy Directive 15 (SPD 15), which has not been revised since 1997 (62 FR 58782:58782-58790), regulates "Standards for the Classification of Federal Data on Race and Ethnicity." It would require a change to SPD 15 to switch from the two-question format to the one-question format.

The summary starts with comparing whether you declare yourself to be Hispanic or Latino, or not Hispanic or Latino. Figure 1 is a controlled comparison between the 2010 and the 2020 Census that makes the point that the percentage of the population that declared in 2020 that they were of Hispanic or Latino origin did increase substantially (all the figure are taken from https://www.census.gov/library/stories/2021/08/improved-race-ethnicity-measures-reveal-united-states-population-much-more-multiracial.html).

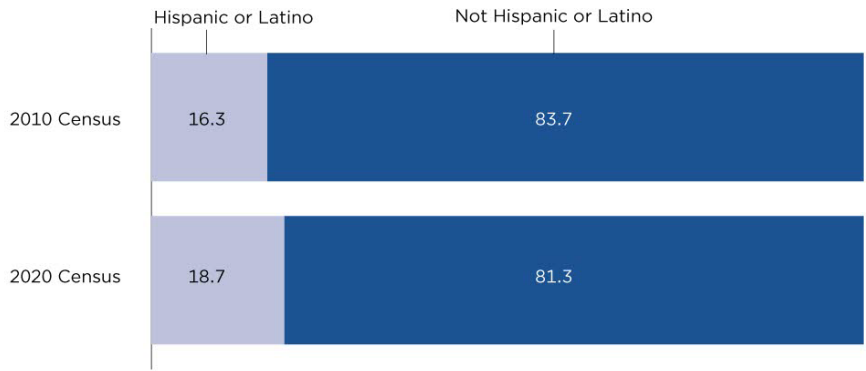

Source: https://www.census.gov/library/stories/2021/08/improved-race-ethnicity-measures-reveal-united-states-population-much-more-multiracial.html

The reason this is somewhat more complicated than one would like is shown in Figure 2. The top panel is the 2010 Census broken down by the three ways in which we categorized race data in the U.S. Because you're allowed to check as many boxes as apply, from six categories. The sixth category, "some other race," is not in Statistical Policy Directive 15. It was added by a Congressional amendment to Title 13-- ahead of the 1990 Census--and has not been repealed.

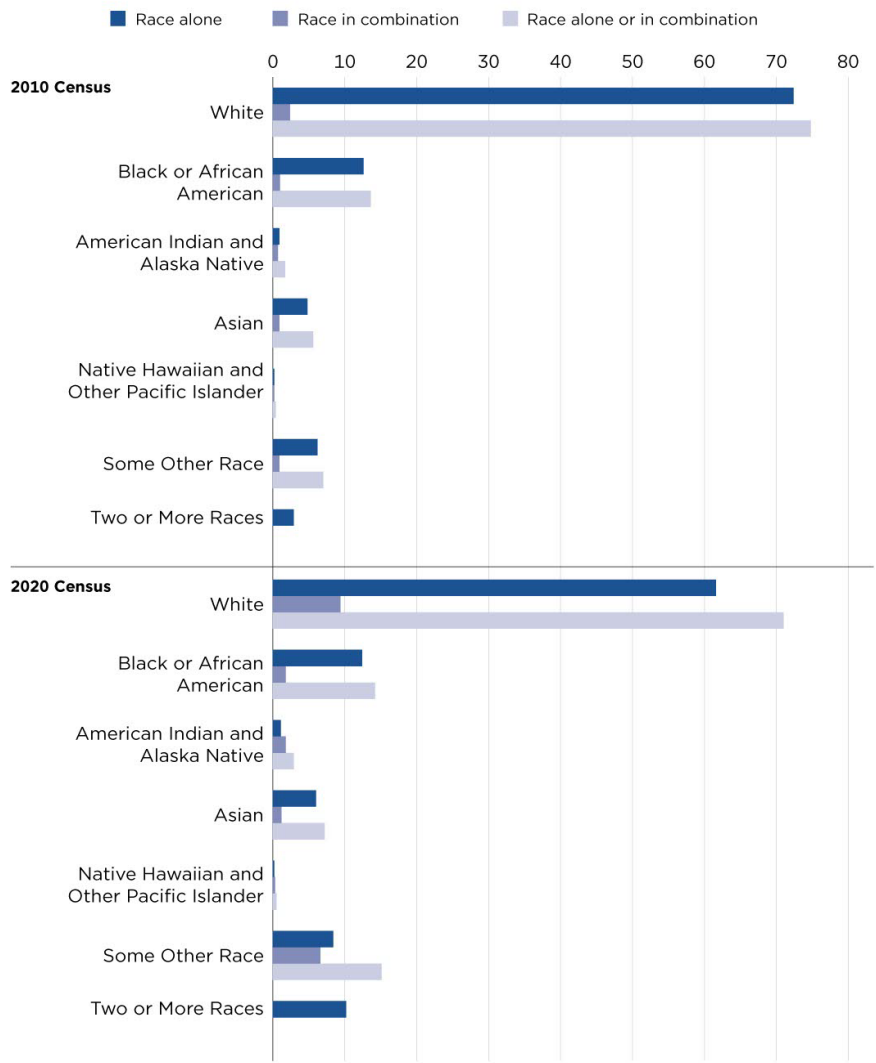

Source: https://www.census.gov/library/stories/2021/08/improved-race-ethnicity-measures-reveal-united-states-population-much-more-multiracial.html

And as you can see, comparing the 2010 Census to the 2020 Census, there have been some fairly substantial changes in the "Race alone or in combination" statistics.

The percentage changes in the race groups between 2010 and 2020 are shown in Figure 3. What really stands out is the large increase in the percentage of the population that reported White in combination with some other race group, and also the very large increase in the declaration of two or more races. It turns out that that's intimately related to the way in which the Census Bureau, by Statistical Policy Directive 15, must separately classify Hispanic origin and race categories.

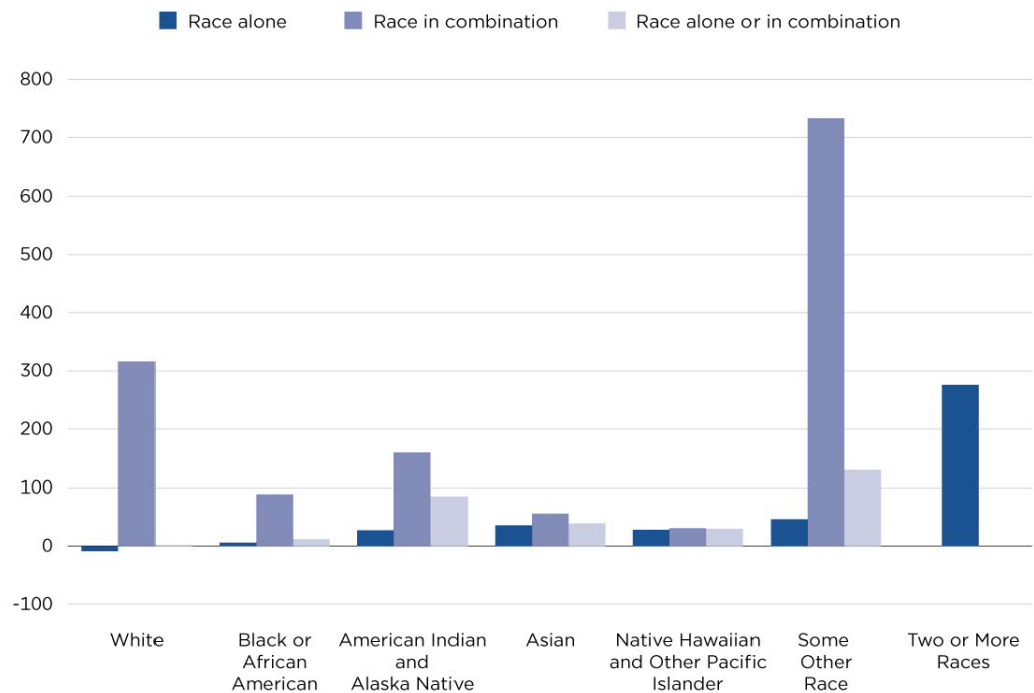

Source: https://www.census.gov/library/stories/2021/08/improved-race-ethnicity-measures-reveal-united-states-population-much-more-multiracial.html

Many people who declare Hispanic origin decline to characterize themselves in one of the five official race categories in SPD 15 and click some other race. Many people write in Hispanic origin as their response in one of the categories of the race question, which the Census Bureau then recodes, by instruction, to some other race in the race category. So that's why you see big differences between 2010 and 2020. The questions matter.

It is this dichotomy between Hispanic race and ethnicity and the official five categories, plus some other race, which Statistical Policy Directive 15 mandates, that creates the difficulty in interpreting race and ethnicity statistics as they evolve over time.

If you want to do well-formed comparisons, you need to understand that there were many improvements made to the question that was asked in the 2020 Census, compared to the one that was asked in 2010 and on the American Community Survey from 2010 up to and including 2019.

The changes to the race question are typically implemented on the American Communities Survey in the same year as the Census. That was also the case in 2020. The demographers at the Census Bureau are confident that the overall racial distribution differences are largely due to improvements in the design of the two-question format, in particular the addition of write-ins under White and under Black and African American allowing the respondent to declare additional information in those categories.

That's one big change. The other big change is that, instead of limiting the data processing to 25 characters, which is what was done in 2010 and in the ACS processing until 2020, the data processing now takes the complete answer. And instead of limiting the coding to at most two output categories, as many as eight can be coded.

There's an excellent webinar ([Collecting and Tabulating Ethnicity and Race Responses in 2020 Census](#)) and blog ([https://www.census.gov/newsroom/blogs/random-samplings/2021/08/improvements-to-2020-census-race-hispanic-origin-question-designs.html](https://www.census.gov/newsroom/blogs/random-samplings/2021/08/improvements-to-2020-census-race-hispanic-origin-question-designs.html)) that go through these changes and show the net effect on the way a question was answered. You have to make comparisons between the two censuses with caution because the form of the questions changed.

Now, to survey specialists in the audience, that the form of the question would affect how you do comparisons to other questions is a tenet of faith and correct. But it's not so obvious in the reporting, because the reporting standards look unchanged.

But the question wasn't asked the same way and it wasn't coded the same way. A little background. The 2015 National Content Test ([https://www2.census.gov/programs-surveys/decennial/2020/program-management/final-analysis-reports/2015nct-race-ethnicity-analysis.pdf](https://www2.census.gov/programs-surveys/decennial/2020/program-management/final-analysis-reports/2015nct-race-ethnicity-analysis.pdf)) remains the most extensive test of alternative ways of asking race and ethnicity of the U.S. population. It was a test involving more than 1,000,000 households, conducted in the lead up to the 2020 Census.

The two-question format was used on the 2020 Census, in spite of that study's conclusion that the single-question format would largely eliminate the problem of some persons of Hispanic origin preferring to declare a racial category rather than an ethnicity category, by eliminating the distinction and just letting them declare as they wish.

The one-question format basically wipes out the "some other race" category for most Hispanic respondents. But it had to be approved by the Office of Management and Budget before the Internet Self-Response Instrument for the 2018 End-to-End test of the 2020 Census was conducted. That instrument was turned on in December of 2017.

The instrument was programmed both ways. Because there was no approval from the Office of Management and Budget to switch the question format, i.e., to revise Statistical Policy Directive 15, the two-question format was turned on. And once it was the tested format, it went forward into the 2020 Census.

Despite the problems raised by the changes, we can do some things with cautious comparisons. I'm going to show you some comparisons that also involve the question change between the 2019 ACS and the experimental one-year 2020 ACS (Table 1). All right. I've highlighted the comparable statistics between the 2010 and 2020 Censuses in their reporting format.

Table 1

Responses to Race Questions

Panel A

| Population | United States (2020 Census) | | United States (2010 Census) | |
|---|---|---|---|---|
| | Count | Percent of Total | Count | Percent of Total |
| Total: | 331,449,281 | | 308,745,538 | |
| White alone | 204,277,273 | 61.6% | 223,553,265 | 72.4% |
| Black or African American alone | 41,104,200 | 12.4% | 38,929,319 | 12.6% |
| American Indian and Alaska Native alone | 3,727,135 | 1.1% | 2,932,248 | 0.9% |
| Asian alone | 19,886,049 | 6.0% | 14,674,252 | 4.8% |
| Native Hawaiian and Other Pacific Islander alone | 689,966 | 0.2% | 540,013 | 0.2% |
| Some Other Race alone | 27,915,715 | 8.4% | 19,107,368 | 6.2% |
| Two or More Races | 33,848,943 | 10.2% | 9,009,073 | 2.9% |

Panel B

| Population | United States (2020 ACS 1-Year Experimental) | | | United States (2019 ACS 1-Year) | | |
|---|---|---|---|---|---|---|
| | Estimate | Percent of Total | Margin of Error | Estimate | Percent of Total | Margin of Error |
| Total: | 329,484,119 | | 13,206 | 328,239,523 | | |
| White alone | 206,619,960 | 62.7% | 163,627 | 236,475,401 | 72.0% | 99,212 |
| Black or African American alone | 39,839,863 | 12.1% | 71,421 | 41,989,671 | 12.8% | 77,381 |
| American Indian and Alaska Native alone | 3,239,492 | 1.0% | 38,164 | 2,847,336 | 0.9% | 33,671 |
| Asian alone | 18,943,940 | 5.7% | 41,975 | 18,636,984 | 5.7% | 39,562 |
| Native Hawaiian and Other Pacific Islander alone | 601,228 | 0.2% | 12,460 | 628,683 | 0.2% | 17,795 |
| Some Other Race alone | 22,316,584 | 6.8% | 176,213 | 16,352,553 | 5.0% | 122,060 |
| Two or More Races | 37,923,052 | 11.5% | 199,358 | 11,308,895 | 3.4% | 119,592 |

Source: John Abowd's calculations.

*Caution: Comparisons between the 2020 and 2010 Census confound changes in the question design with changes in the underlying demographic characteristics. Comparisons between the 2020 Experimental 1-year ACS and the 2019 1-year ACS confound changes in the question design, pandemic-related nonresponse to the ACS and the underlying demographic characteristics.*

But they are not comparable in the way the question was asked. That big change in White alone that we saw in Figure 3-–it went way down--was largely because there were more opportunities for people who self-respond as White to also supply additional answers.

Now look at panel B, that compares the 2019 ACS, which used the old question format and the old coding, to the 2020 experimental one-year ACS. It is tempting, and not entirely incorrect, to conclude that most of that difference is due to the change in the format and the coding. But there are some major qualifications.

The first is that some of that difference is real and it's going to take a very detailed study, which Census Bureau demographers are undertaking, to try to untangle that using their database of basically every race and ethnicity response that has ever been supplied to the Census Bureau in written form.

That database is used to generate the coding algorithms for both the redistricting data and for the detailed race data that will be published later. In addition, there was a serious non-ignorable missing data problem with the 2020 ACS that was partially addressed in order to publish the 2020 one-year ACS tables.

They were only published at higher levels of geography. They were clearly labeled as experimental because that fix was implemented very quickly. It's well-documented on the Census.gov site (See Rothbaum et al. 2021 at [Addressing Nonresponse Bias in the ACS Using Administrative Data (census.gov)](#)).  A much more extensive fix has been implemented for the five-year data product that was released last week.

The experimental tag has been removed but further research is still required to address the non-ignorable missing data, especially for-- questions like race and ethnicity. Thank you for your patience.

SONYA RAVINDRANATH WADDELL:

Thank you, John. I am going to turn it over to you, Mark to talk about what we know about how people self-identify. I know you've done work in this space.

MARK HUGO LOPEZ[1]:

Thank you. John, that was really great. I think it's important to keep in mind how the Census Bureau asks about race and ethnicity. As we've been talking about, as Sonya mentioned, this is one of the key ways that we, as researchers actually think about race and ethnicity.

But our work at the Pew Research Center has shown that there are other ways in which the public thinks about their identity. The Census forms, and Census-style questions, is just one way that one might actually ask somebody about how they think about themselves.

Understanding that sense of identity can vary a lot, depending on the context in which the question is asked: who's asking you, where is it being asked, at what time in your life is it being asked, all these things can have an influence on how somebody might respond.

---

[1] Nothing in these remarks is to be seen as attributing a particular policy or lobbying objective or opinion to Pew Research Center, or as a Pew Research Center endorsement of a cause, candidate, issue, party, product, business, organization, religion or viewpoint.

for example, from the questions that Census Bureau has, particularly when it comes to Hispanics and Latinos, race seem to suggest that many folks identify as some other race. They'll oftentimes write in something like, "Mexican" or "Latin American" or "Hispanic". In fact, in 2010, most of the responses were one of those three-- write-ins in the "some other race" category as an example.

Whereas at Pew Research we've been trying to go in a somewhat different direction. When you ask somebody, "Are you indigenous?" Or "Do you have indigenous roots?" You'll find that maybe about a quarter of Latino adults will tell you, yes, they do. What about if you ask, "Are you Mestizo or Mestiza, mixed in some way?"

You'll find that maybe about a third of Latino adults will say that. And what about, "Do you consider yourself Afro-Latino or Afro-Cubano, or Afro-Dominican?" You might find that maybe anywhere from 12 to 25%, depending on the survey, might say that that's what they consider as part of their identity.

Why am I talking about all these different possibilities? Because some of these things may not necessarily be obvious in the Census Bureau question. And people may not know exactly how to report this or identify themselves in Census forms.

And that's why you see many Hispanics in the Census form will say in the first question, "I am Hispanic or

Latino." And they'll write in their origin. And then in the race question, they'll say, "I'm some other race," and write in their origin again. So you get Mexican-Hispanic-Mexican, and then Mexican is my race.

That to me is really interesting. There's a couple of ways to think about it. On the one hand, maybe people are confused by the form. I'm not sure that that's right. I don't think people are confused about their identity. But maybe the form could be clearer.

Or it could be that people are trying to tell you, "My race is-- I think of my racial identity as Mexican." And that's something that we hear often in our work at the Pew Research Center. We've been trying to explore identity in other ways as well for other racial and ethnic groups and also for immigrant groups.

When we talk about racial identity, I think that we oftentimes have to try to think outside of the box that we're usually thinking about when it comes to the Census Bureau's forms. I will say, though, I don't know what the right answers are. I do think that more research is needed. And I'm excited that the Census Bureau is doing some really great research on how Americans think about and report their racial identities.

SONYA RAVINDRANATH WADDELL:

I will say that, with an Indian father and a white American mother, I have filled out all sorts of things on Census forms that I don't want to report here.

Camille, let me turn it over to you.

CAMILLE BUSETTE:

A couple things that I wanted to mention relevant to John's excellent presentation and Mark's expansion of that.

One of the interesting things about this whole discussion around race, and particularly something that Mark had mentioned, is that race is a very American kind of concept.

As we know, it may not be something that is actually real. It's a construct. It doesn't surprise me that people would say, "My race is Mexican." Unless you've been here a really long time and have really kind of absorbed the construct of race as it used in the U.S. you would. It's an odd and awkward kind of concept.

So it doesn't surprise me that people would say, "I'm Dominican," or "I'm Mexican," *et cetera*. But when you think about race biologically, it doesn't really exist.

I think one of the interesting things about John's presentation and the work going on in the Census is that it is starting to try to capture what-- in a sense-- is a really multi-racial society., trying to in some ways push the boundaries of the strictures of race and identity that we have in the U.S. I think it's trying to get us closer to reality. I really applaud that. There are a couple things I want to say that are a little bit removed from this particular discussion about identity and the categories in the Census.

What is interesting about Census data collection is that it drives a lot of our policy conversations. When you're thinking about the ways in which particularly federal public policy is made, a lot of times the categories, the concepts, the framing of public policy, is based on categories and framing that come out of data collection.

Public policy can be driven by what is collected. There are a lot of things that aren't collected. What we're talking about today here is some of the things that could be collected or could be shaped that either are showing up early in some of the sort of test studies that the Census does, or aren't showing up at all, that could change the set of options that are available in public policy and the kinds of questions we ask in public policy.

That's why it's really important to track what's happening in the Census and to think about the debate, and to think about how we influence that. Otherwise, if we don't do that, where we have missing data so to speak, or where we don't have the words and the framing for concepts and experiences that exist in the U.S. public policy space, is then, in those cases, shaped by public opinion, which is shaped by assumptions, misconceptions, all sorts of things.

That's why it's really important for us to think about what's in, what's out, and how are we going to adjust the things that aren't in there and aren't conceptualized in a way that is part of the American experience.

SONYA RAVINDRANATH WADDELL

As we think about economic outcomes, what are we measuring? What do we need to be measuring? What if you could wave a magic wand and not just pick the economic outcomes that we're measuring, but also what sort of classifications we need to have to be able to make better policy

MARK HUGO LOPEZ:

I'm going to answer a little bit to the immigration question because this is something I think that's very important. One of the things that of course happened over the last 50 years is that we've had over 59 million people arrive from outside the United States to come and live and work here.

Of course, not everybody stayed, and some have passed away. We have about 45 million immigrants in the country today. What's important about that is that immigrants and their U.S.-born children have been the key drivers of U.S. population growth over the course of the last five decades.

They'll continue to be so into the next coming decades. In fact, to give you some numbers, about 85% or more of the nation's population growth has been because of immigrants arriving here, or their children being born here, or the offspring of those children being born here.

It's also a big number, the biggest that the country has ever seen. We're certainly approaching record levels of the share of the population that's foreign born.

What should we be measuring, though? I have long found that in our work on populations that have large groups of immigrants that it's important to know not only if somebody is born in another country, but to also know if the children are the first generation born here, so are the U.S.-born children of immigrant parents. And then if somebody is U.S.-born to U.S.-born parents, they have some sense of immigrant generations.

We do get this in some Census Bureau and some U.S. government data sources. For example, the Current Population Survey does collect birthplace of parents and the birthplace, of course, of the survey respondent. You can actually calculate this and get this sense of generations. The American Community Survey does not.

It would be interesting to be able to have something like this on the American Community Survey to, again, provide us with some good estimates of what we know about the immigrant generation experience. If we wanted to go a little farther, you could ask about the birthplace of grandparents. That would be something else that you could (that might be a lot to ask for).

But that is another thing that we have tried at the Pew Research Center, and found it's been successful at getting a sense of who might even be what is fourth or higher generation. And there is a distinct economic and social story for those folks compared to third, second, and immigrant generations.

Camille Busette:

What's interesting about Census data is that it allows us to get counts, and percentages, and generational variations, *et cetera*. What's interesting about outcomes is sometimes you don't understand why it is that we see some of the differences that we see. I'm just going to give you an example. And I know John's going to have something to say about this.

When we think about wage gaps and we think about unemployment gaps in the U.S. one of the things we know is that women earn less than men. We know that. We also know that when we look at unemployment data, for instance, that Black men always have a much higher unemployment rate than white men.

So those are just facts. But what it doesn't allow us to understand is how hard it is for women to work for 83 cents on the dollar if you're white, 75 cents on the dollar if you're Black, 69 cents on the dollar if you're Latina, right, and how hard your life really is.

Same thing for Black men. Black men might have to put in 20 different applications for a job before they get a job. What those data points don't capture is the level of difficulty that so many people in this country experience in trying to get to a point where they are actually doing well economically and financially.

What's interesting is when you are a consumer of that data, you have to interpret that data. The Census can't interpret the data for you. But somebody who is a Black man and 40 looking for employment is going to have a totally different experience than a white man-- all other things being equal-- looking for employment.

As you're thinking about how you report on that it's not just saying, "Oh. Well, you know, Black men have a higher level of unemployment." We have to think about how we fill that space that the data cannot show us, and we can't expect the Census to fill in for us.

JOHN ABOWD:

Well, I was invited to comment but I don't have any disagreement with anything that you said. I will say though that a supporting argument to what you've said is that social network studies have shown that the most important factor in getting a job is who you know and those referrals. And so, separation of your friend network, or separation of your family network is an important outcome that we need to take into account in trying to address the racial and ethnic disparities that you so nicely summarized.

SONYA RAVINDRANATH WADDELL:

One of the things that we did talk about in the planning call was, how we think about wealth and income, and wages. How should we think about economic outcomes across our populations?

JOHN ABOWD:

I'll start but I have to say that that we're in an information-gathering mode for trying to understand what additional statistical products would help address many of the concerns that you raise.

I will say that we are investigating ways to harvest what could be described as the low-hanging fruit. We don't know very much about the racial and ethnic backgrounds of business owners and of the kind of people who develop businesses.

We have been trying to augment that data collection. The economic programs at the Census Bureau have been looking to supplement their demographic data. Other statistical and non-statistical agencies have come to the Census Bureau and asked us to help with supplying them with race and ethnicity data deliberately coded so that it's comparable with the other products that the Census Bureau has released. We're trying to help in that. But I'd love to hear more ideas.

CAMILLE BUSETTE:

I'll comment on your comment. We like to get, as researchers and policy people, disaggregation to the nth degree, which, of course, is an impossible ask for the Census.

I have seen some interesting studies of questionable integrity in the following sense, that have claimed that-- there are lots and lots--a very significant portion of the Black population, has started to become upper class. There's some attributes to that, like home ownership and the type of job and education one has.  It's true that there has been obviously a lot of progress and accomplishment within the category of African-Americans in the U.S. over the last 50 years. But what I think is interesting about those kinds of studies is they don't disaggregate by the type of African-American.

I think about this because my parents are also immigrants. You have what I would call indigenous African-Americans, these are the descendants of Africans who were enslaved here in the U.S. And then you have people from, for instance, the Caribbean who are descendants of Africans who were enslaved in other places.

And then you have people who come here as immigrants from the African continent. Those are really different experiences, as Mark has mentioned. And they lead to, I think, different kinds of outcomes.

They lead to different educational outcomes, different professional outcomes, et cetera. So when you don't disaggregate, you can get studies that say, "Oh. You know, African-Americans are doing really well." And I bet if you look at that you might find that a good proportion of those people came from "immigrant"--and I'm going to use this as sort of a quotation--backgrounds, but were highly placed in the places that they came from.

They just kind of found the keys to replicate that here. And then there are indigenous African-Americans who are obviously are very, very accomplished as well. But I think, particularly when you're thinking about policy and politics, it's important to get that disaggregation so that the story we're telling about can be as nuanced as the actual experiences themselves.

MARK HUGO LOPEZ:

If I could add to that. I think it's important to capture peoples' identities in so many different ways. That sort of disaggregation is something that we've been doing for a long time at Pew Research on the Latino side. We just started doing more work looking at Black Americans. We've done two large surveys that allow us to tell the differential story across the population.

We're trying to change our editorial frame as well, to tell a story about not just about Black Americans, but the differences within the population. So we can say something about immigrants who might trace their roots to Africa versus immigrants who might trace their roots to the Caribbean, versus say people who say that they're Black but also another race, so more than one race, or people who are Black and Hispanic, or even interestingly, it's actually quite hard to do this in survey data, Black Americans by different age groups, 18 to 29, 30 to 44, 45 to 64, and 65 and older.

It's actually quite hard to be able to have enough data, particularly because young people don't want to participate in surveys. So, please, if you're younger, please participate in surveys if Pew Research calls you.

It's important to capture these differences, because there are unique stories depending on the measure. Even knowledge of a family's history and whether or not people know that their ancestors were enslaved--not everybody does. It's interesting to see the differential story that exists within the population. Look for more work from us. We already have some of those data files already publicly available. If you're interested in it, I can point you to one big one from a couple years ago.

AUDIENCE MEMBER:

How can the gender conversation learn from what's been worked on from the race perspective?

JOHN ABOWD:

Thank you for that question. I'm going to take the opportunity to make a technical answer because it's very important, especially in the sexual orientation, gender identity, context that you posed your question in, which is that the vast majority of the surveys that the Census Bureau conducts at the household level rely on a single respondent to respond for all the members of the household.

That is a particularly difficult situation for collecting accurate SOGI (sexual orientation and gender information) information. My former colleague, Nancy Bates, has spent much of her professional career researching it, and is now a part of a National Academy's Consensus Panel, that just released its findings and recommendations. (See: [Measuring Sex Gender Identity and Sexual Orientation for the National Institutes of Health | National Academies](#).) I remember that when we were trying to understand our options, the Census Bureau participated in funding a pilot study.

They designed it to be able to measure the accuracy of single respondent versus self-responding on those questions. That does guide a lot of the design of questions of that sort at the Census Bureau.

Now there are some surveys that we conduct that are self-response only. Some of them are client surveys for the National Center for Health Statistics, which has asked SOGI questions in surveys that the Census Bureau conducted for them.

CAMILLE BUSETTE:

And I will just say amen to that, being North African and not feeling like the categories fit. They don't. There are just so many populations for whom that is true. I'm sure the Census is grappling with those issues.

JOHN ABOWD:

The Middle Eastern, North African classification was in that 2015 Census test and was a part of the two-

question format as well as the one-question format. But it was not a part of the approved two-question format that was used in the 2020 Census because it also requires a change to Statistical Policy Directive 15.

AUDIENCE MEMBER:

I'm coming at this from a practitioner perspective, as an investor. Many years ago I managed a pool of capital from a pension fund. And the pool of capital was a private equity fund with the mandate to invest in minority- and women-owned businesses in California, businesses in the inner cities, businesses in the rural areas of California.

And in my bench-marking, there wasn't enough data that got at the very specific level of, let's just say, productivity figures for Latin-owned businesses, or minority-owned businesses. The dream database that I could've had would be, like, a balance sheet-- income statement and a cash flow statement broken down by all these socioeconomic—for all these identity variables.

The identity variables with business outcomes is, I should say, just lacking. Now obviously there are public policy implications. But even in the private sector, that's a rich database that would just transform so much and be able to deploy capital in areas and markets that previously have just been overlooked by institutional investors on Wall Street.

John, if you're looking for ideas, I used to read a lot of those PUMS (Public Use Micro Sample) databases from

back-- 20, 30 years ago. If we had the equivalent markers for the economic Census that comes out every five years, particularly for productivity numbers and any of the balance sheet figures on profitability and such, that would be heavenly in terms of moving capital into those underserved markets that I think we care about.

CAMILLE BUSETTE:

Well, John's writing so I think that's a good sign. Thank you for that suggestion. That data is very hard to collect. Absent some kind of federal directive to actually collect it, the way in which you get that is either through other surveys that have been done-- sometimes by the Fed or individual Fed banks. Or, banks will have that information if somebody has applied for a loan, but that would not be providing a national database. It's very, very hard to get that information.

I wanted to go back to your question, Sonya, about wealth, and data around wealth versus income. One of the things that strikes me, particularly in the current environment--are we post-COVID, dealing with COVID or the after-effects of COVID?-- is the question of jobs with benefits. But there is a particular importance of jobs with benefits to wealth creation.

I think that's an important thing to be able to track over time and to be able to be able to connect with race and identity in the way the Census does it. Some of that is tracked in other kinds of federal studies. But it is not consistently tracked alongside the Census tracking.

To do that you'd have to merge databases, which would be tough. But jobs with benefits make it much easier to save. You have fewer expenses that are just coming straight out of your pocket all the time. There's a very rich body of research which shows that part of the racial wealth divide in this country (we use the word racial, but it also applies to Latinos as well) is driven by the fact that some people, a lot of white households, actually have jobs with benefits and have had that over the last 70 to 80 years, versus African-American, Latino, Native American, *et cetera*, households that don't have access to as many jobs with benefits.

SONYA RAVINDRANATH WADDELL:

That's a really good point. Because I can imagine that you would have a sort of a safety net there if you were to have health issues, which is one of the reasons why a lot of households run into trouble.

PATRICK JANKOWSKI:

My name's Patrick Jankowski. I work for an economic development organization in Houston. One of the things we're trying to do is launch a DI initiative, trying to do what we can in the community. One thing we get frustrated with is we can find unemployment levels at the national level based on race and ethnicity or educational attainment and gender. But we can't get that data at the local level-- only when it comes out with the ACS. Is there any potential to start reporting unemployment rates by these categories at levels lower than the national level?

JOHN ABOWD:

I'm going to answer that question. But I am not going to speak for the Bureau of Labor Statistics, which is the supplier of local unemployment statistics. I will say that the method of collecting data by direct survey is prohibitively costly for the kind of statistics that you're looking for.

Most of the answers seem to come from administrative data. The Census Bureau has several administrative-data products that you might be able to use. Because they're based on administrative data, they're not as timely as monthly unemployment statistics. But they will give you employment-population ratios and other things.

The one that we publish annually, called "[OnTheMap](OnTheMap)," goes all the way down to the block. But it's modeled from the tract down to the block. It's not a direct measurement.

CAMILLE BUSETTE:

I'll just say that we've actually looked at that kind of data. And what we have found that if you want that level of data, it's usually because the county or the city actually has spent resources on collecting it.

It's hard to get it from the Census, for instance. One of the things I might suggest is as we have ARP money flowing, that you might think about asking, your local jurisdiction, either the county or city, whichever makes more sense, to start collecting that data.

It could be a moderate to heavy lift-- and a moderate to heavy investment. But we do have dollars flowing now for infrastructure and other kinds of investments like that. Something like this could qualify for that. But in my experience, you literally have to hope that some county or city decided to put their resources there.

SONYA RAVINDRANATH WADDELL:

I can add to that. All of the Federal Reserve Banks have a small business credit survey. But the granularity and the timeliness is another question.

We've found that some of our jurisdictions actually collect that data already, but don't have the analytical capability to do anything with it, or to check the data. Sometimes we've had luck, on particular projects going to the city or to the county and getting that data from them and provide information back to them which can be sort of a mutually beneficial arrangement.

But, even for us in the Fifth Federal Reserve District, we can't do that across the board.

# Biographies

Sonya Ravindranath Waddell is a vice president and economist at the Federal Reserve Bank of Richmond. Waddell has responsibility for the Regional and Community Development research areas within the Research Department, including setting strategic direction for various data products, surveys, and other regional and local analysis. In addition, she directs the incorporation of regional information into FOMC policy preparation for the Richmond Fed. Her work involves analyzing economic trends, writing for a variety of publications, and presenting on regional and national economic conditions. Prior to joining the Richmond Fed in 2008, Waddell worked as an economist in the Virginia Department of Planning and Budget and at ICF International in Washington, D.C. She earned her bachelor's degree from Williams College in 2001 and her master's degree from the University of Wisconsin-Madison in 2006.

John M. Abowd is the U.S. Census Bureau's associate director for research and methodology, and chief scientist. He was named to the position in June 2016. The Research and Methodology Directorate leads critical work to modernize our operations and products. He is currently leading the agency's efforts to create a differentially private disclosure avoidance system for the 2020 Census and future data products. His long association with the Census Bureau began in 1998 when he joined the team that helped found the longitudinal employer-household dynamics program. In 2008, he led the team that created the world's first application of a differentially private data protection system for the program's OnTheMap job location tool. Abowd is also the Edmund Ezra Day Professor emeritus of economics, statistics, and data science at Cornell University. He is a fellow and past president of the Society of Labor Economists. He is also a fellow of the American Association for the Advancement of Science, American Statistical Association, and Econometric Society, as well as an elected member of the International Statistical Institute, and was the 2016 recipient of the Julius Shishkin Award for Economic Statistics. He earned his M.A. and Ph.D. in economics from the University of Chicago and A.B. in economics from the University of Notre Dame.

Camille Busette is director of the Brookings Race, Prosperity, and Inclusion Initiative and a senior fellow in Governance Studies, with affiliated appointments in Economic Studies and Metropolitan Policy. Busette has dedicated her career to expanding financial opportunities for low-income populations. She has previously held roles at the Consultative Group to Assist the Poor (CGAP), the Consumer Financial Protection Bureau, the Center for American Progress, and at EARN, a provider of micro savings services to low-income families in the U.S. Her private sector experience includes roles as the Deputy Director of Government Relations for PayPal, the Head of the Consumer Data Privacy function at Intuit, and the Director of the Consumer and Market Research division at NextCard.

Dr. Mark Hugo Lopez is director of race and ethnicity research at Pew Research Center, a nonpartisan fact tank that informs the public about the issues, attitudes and trends shaping the world. It does not take policy positions. The Center is a subsidiary of The Pew Charitable Trusts, its primary funder. Lopez leads planning of the Center's research agenda focused on chronicling the diverse, ever-changing racial and ethnic landscape of the United States. He is an expert on issues of racial and ethnic identity, Latino politics and culture, the U.S. Hispanic and Asian American populations, global and domestic immigration, and the U.S. demographic landscape. Lopez was previously the Center's director of Global Migration and Demography, and of Hispanic research. Lopez is the co-editor of *Adjusting to a World in Motion: Trends in Global Migration and Migration Policy*. He is a co-author of *The Future of the First Amendment* and has contributed chapters to several books about voting and young Latinos. Lopez received his doctorate in economics from Princeton University. He is an author of reports about the Hispanic electorate, Hispanic identity and immigration.